\documentstyle[aps,prb,epsf]{revtex}
\newcommand{\bm}[1]{\mbox{\bf{#1}}}
\begin{document}
\draft
\title{{\em{Ab-initio}} prediction of the electronic
and optical excitations in polythiophene:
isolated chains versus bulk polymer} 
\author{J.-W.\ van der Horst, P.A.\ Bobbert, P.H.L.\ de Jong, M.A.J.\ Michels}
\address{
Dept.\ of Applied Physics, COBRA Research School \& Dutch Polymer Institute, Technische 
Universiteit Eindhoven, P.O.\ Box 513, NL-5600 MB Eindhoven, The Netherlands}
\author{ G.\ Brocks, P.J.\ Kelly}
\address{Computational Materials Science,
Faculty of Applied Physics, University of Twente, P.O.\ Box 217, NL-7500 AE Enschede,
The Netherlands\\}
\maketitle
\begin{center}(\today)\end{center}

\begin{abstract}
We calculate the electronic and optical excitations of polythiophene
using the $GW$ approximation
for the electronic self-energy, and include excitonic effects by solving the electron-hole
Bethe-Salpeter equation. Two different situations are studied: excitations on isolated
chains and excitations on chains in crystalline polythiophene.
The dielectric tensor for the crystalline situation is obtained by modeling the
polymer chains as polarizable line objects, with a long-wavelength polarizability tensor obtained
from the {\em{ab-initio}} polarizability function of the isolated chain.
With this model dielectric tensor we construct a screened interaction
for the crystalline case, including both intra- and interchain screening.
In the crystalline situation both the quasi-particle
band gap and the exciton binding energies are drastically reduced in comparison with 
the isolated chain. However, the optical gap is hardly affected. 
We expect this result to be relevant for conjugated polymers in general.
\end{abstract}
\pacs{78.40.Me,71.20.Rv,42.70.Jk,36.20Kd}

\section{{Introduction}} 
Semiconducting conjugated organic polymers have received increasing interest
in recent years, especially since the discovery of electroluminescence \cite{ppv} of these materials. 
The charge carriers and excitations in these materials have been
studied extensively both experimentally and theoretically, but many important
fundamental issues still remain unsolved. For instance, the magnitude of the exciton binding energy 
in these materials is still disputed.\cite{bredasheeger} This is a very important quantity,
since e.g.\ in photovoltaic devices
(solar cells) one would like to have a small binding energy, which facilitates the fast
separation of charges, while in electroluminescent devices
such as LEDs a larger exciton binding energy, to increase the probability of fast 
(radiative) annihilation of electron-hole pairs, is desirable.

In conventional semiconductors such as Si and GaAs
the optical excitations are well described in terms of very weakly bound electron-hole pairs
(so-called Wannier excitons) with a binding energy of the order of 0.01 eV. 
In crystals made of small organic molecules such as anthracene, the exciton 
is essentially confined to a single 
molecule (Frenkel exciton), leading to a binding energy of the order of 1 eV.
The question is where exactly conjugated polymers fit in between conventional 
semiconductors on the one hand and molecular crystals on the other:
negligibly small (0.1 eV or less~\cite{onetenthorless}), intermediate 
($\sim$ 0.5 eV~\cite{intermediate}), 
and large ($\sim$ 1.0 eV~\cite{oneev,melppp,wohl}) binding energies have been proposed. 

{\it{Ab-initio}} calculations, on a variety of conjugated polymers, within 
the Local Density Approximation of
Density Functional Theory (DFT-LDA) yield equilibrium structures in very good 
agreement with experiment.\cite{vc,bkc,gdc,abd} The Kohn-Sham
gaps in these calculations are typically 40\% smaller than the optical
band gap (absorption gap).
In cases where calculations for both isolated chains and the
crystalline situation were performed, small differences (0.1 to 0.2 eV for
gaps of $\sim 3.0$~eV) in Kohn-Sham band gaps were found.\cite{gdc,abd}
However, it is well known that the Kohn-Sham
eigenvalues formally cannot be interpreted as excited state energies.\cite{godbydeltaverhaal}
Moreover, excitonic effects are not taken into account in these calculations. 

An {\em{ab-initio}} many-body calculation within the $GW$ Approximation~\cite{gw} ($GW$A)
was performed for poly-acetylene (PA) by Ethridge {\em{et al}}.\cite{ethridge}
They claim that their quasi-particle (QP) gap, excluding excitonic effects,
is in agreement with the experimental
absorption gap. 
This result seems to be in contrast with a more recent
calculation by Rohlfing and Louie \cite{oeicp} of both one- (QP) and two-particle (exciton)
excitation energies for PA and poly-phenylene-vinylene (PPV) chains.
Their absorption gaps are in good agreement with experiments,
but the inclusion of excitonic effects proves to be crucial for this.
However, their exciton binding energy of 0.9 eV for PPV is much larger than recently
obtained experimental
values: $0.35 \pm 0.15$~eV for an alkoxy-substituted PPV,~\cite{intermediate}
and $0.48\pm 0.14$~eV\cite{alvpriv} for unsubstituted PPV.

\begin{figure}
\leavevmode
\centerline{
\epsfysize10cm
\epsffile{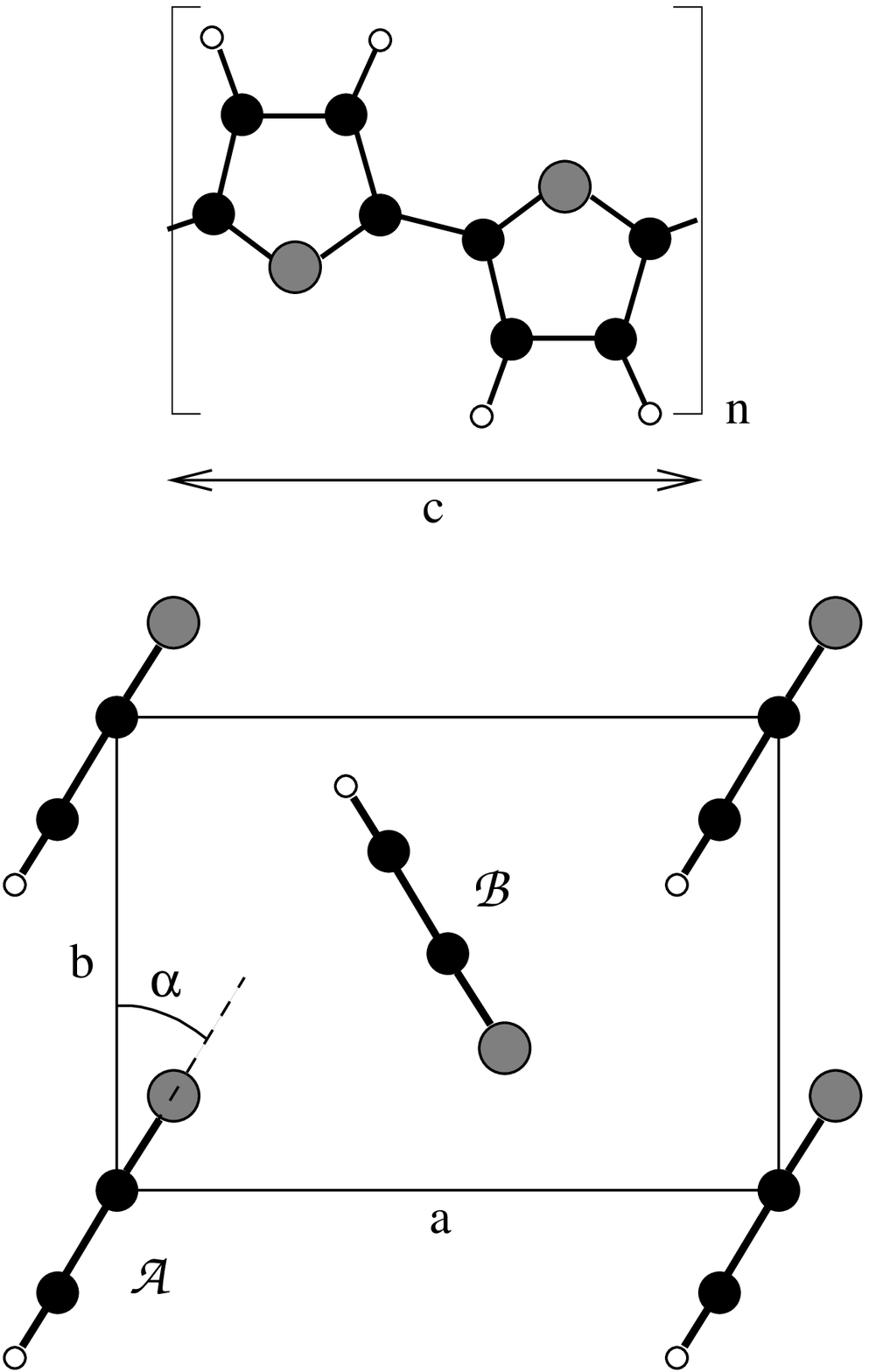}}
\caption{Monomer and crystal structure of polythiophene. Top panel: A single chain of polythiophene.
Black atoms are carbon, gray sulfur, white hydrogen. Bottom panel: crystal structure 
(from Ref.\ \protect\onlinecite{xtal}) as seen
perpendicular to the chain direction ($a$ = 10.5 a.u., $b$ = 14.5 a.u., $c$ = 14.8 a.u.,
$\alpha = 31.2^\circ$).} 
\label{xtal2}
\end{figure}

In a recent Letter,~\cite{onszelf} hereafter referred to as I, 
we focused on the differences in excitations
between an isolated polythiophene (PT) chain, see Fig.\ \ref{xtal2}, and crystalline polythiophene.
For the isolated chain, we found an absorption gap in good agreement with 
experiment, but the energy differences between the various exciton levels were too 
large.~\cite{sakurai}
After including the screening by the surrounding
chains, both the optical gap and the exciton transition energies were in good agreement
with the experimental values. 
The difference in screening between an isolated polymer chain and a condensed polymer medium
can be explained as follows.
For an isolated quasi-one dimensional system, such as a single polymer chain in vacuum,
there is no long-range screening.~\cite{herman,brsawa,schulz2}
A way to understand this is to realize that if we have two charges
on a polymer chain with a separation larger than the width of the chain ($\sim
7$ a.u.), most field lines connecting the charges will be outside the
chain.
If the chain is embedded in a medium (possibly, but not necessarily,
consisting of similar chains), the medium will provide a long-range screening
of the Coulomb interaction.
The screened Coulomb interaction determines both the QP energies (including
the band gap) and the exciton energies. 
Long-range screening reduces both the QP band gap and
the exciton binding energies. Apparently, there is a near cancellation between
the change of the QP gap and the exciton binding energy, meaning that the optical
absorption gap, which is the difference between the two, is influenced much less by
introducing screening.			

In I, a 
method for the calculation of the dielectric tensor of crystalline 
polythiophene from the {\em{ab-initio}} single chain polarizability function was introduced, without giving
any details. Further, some novel technical procedures in the $GW$A calculation were used,
in particular in the handling of the Coulomb divergence both in real and reciprocal space.
Details of these approaches, as well as of the calculation of the
quasi-particle energies and
exciton binding energies, will be explained here.
The paper is organized as follows:
in the next section (\ref{compmeth}) we explain
the computational methods employed to calculate the quasi-particle bandstructure, to
regularize Coulomb interaction, to calculate the
exciton binding energies and the dielectric tensor. In Section~\ref{Results},
we will present results for the electronic and optical excitations of
both the isolated chain and bulk PT. In section~\ref{Discuss} we will discuss these
results and compare them to other calculations, and draw our conclusions.

\section{{Computational methods}} \label{compmeth}
Many successful {\em{ab-initio}} calculations of the QP band structure of conventional
anorganic semiconductors have been performed within the $GW$A~\cite{gw} 
for the electronic self-energy $\Sigma$ of the one-particle Green function.
Very recently, progress has been made in the evaluation of the two-particle Green 
function,~\cite{ben,ar,rl}
from which the optical properties can be obtained. This is done by solving the
Bethe-Salpeter equation\cite{shamrice,strinati} (BSE), which can be mapped onto a two-body 
Schr\"odinger equation for an electron and a hole forming an exciton. We will use these approaches
to calculate the QP band structure and exciton binding energies of PT. The calculational scheme is
as follows: first we perform a DFT-LDA-based Car-Parrinello
calculation, from which we obtain atomic positions,
wave functions and ground-state
energies. We use these as input for the $GW$A calculation, which yields the
QP excitation energies. With the DFT-LDA wave functions and the QP energies we 
calculate the two-particle excitations by solving the BSE. 
This scheme is first applied to the single chain and next to the crystal.
We assume the same atomic geometry for the ground and excited states,
i.e.\ the coupling between electronic and lattice degrees of freedom is neglected.
Experimental data \cite{sakurai} indicate that
energy shifts due to lattice relaxations are of the order of 0.1 eV in PT.
DFT-LDA calculations~\cite{polaron}
predict a hole-polaron relaxation energy of 0.04 eV for 16T ($n$T is an
oligomer consisting of $n$ thiophene-rings). A similar
calculation predicts a triplet exciton relaxation energy of 0.2 eV for 12T.~\cite{tripexc}
Singlet relaxation energies are typically smaller. These values, calculated for oligomers, are
upper bounds for the values in the polymer, since in the oligomers the excitation is
confined, leading to a larger local deviation from the ground state density
and hence to a larger relaxation energy than in the polymer.

\subsection{{The quasi-particle equation}}\label{GWimpl}
We start our calculations with a pseudo-potential plane-wave DFT-LDA calculation~\cite{bkc}
of a geometry-relaxed PT chain in a tetragonal supercell. The plane wave cut-off energy is
40~Ry.
 The length in the chain direction $a_x$ was optimized
and found to be $a_x = 14.80$~a.u.\ (experimental values range from $14.65$~a.u.\cite{xtal} to
$15.18$~a.u.\cite{mo}). In the perpendicular directions
we found that a separation of $a_y = a_z = 15.0$~a.u.\ is enough to consider the chains 
in the DFT-LDA calculation as non-interacting.
The two rings in the unit cell are found to be co-planar and we choose them
in the $y=z$ plane.
We use Hartree atomic units (with the Bohr radius $a_0$ as unit of length and the Hartree as unit
of energy) throughout this article, unless specified otherwise.
The one-particle excitation energies are evaluated by solving the QP equation:

\begin{equation}
\left[-\frac{\nabla^2}{2}+V_{H}(\bm{r}) \right]\phi_{nk}(\bm{r}) +
 \int \left[V_{PP}(\bm{r}, \bm{r}') + 
\Sigma(\bm{r}, \bm{r}', E_{nk}) \right]
\phi_{nk}(\bm{r}')d^3r'
 = E_{nk} \phi_{nk}(\bm{r}),\label{qpeq}
\end{equation}
where $V_H$ is the Hartree potential, $V_{PP}$ the non-local pseudo-potential of the atomic core,
and $\Sigma$ the electronic self-energy. Since in practice the DFT-LDA wave functions and
the QP wave functions are almost identical, we use the former in all calculations.
In DFT-LDA, $\Sigma$ is approximated by:
\begin{equation}
\Sigma(\bm{r}, \bm{r}', \omega) = V_{xc}(\bm{r})\delta(\bm{r}-\bm{r}'), 
\end{equation}
where $V_{xc}$ is the exchange-correlation potential of the homogeneous electron gas.
In the $GW$A, $\Sigma$ is approximated by the first term of the many-body expansion
in terms of the one-particle 
Green function $G$ and the screened Coulomb interaction $W$ of the system.\cite{gw}
In order to calculate $\Sigma$, we follow the real-space imaginary-time formulation of the $GW$A of 
Rojas {\em{et al.}},\cite{spacetime} in its mixed-space formulation.~\cite{msform}
In this formulation, we transform non-local functions $F(\bm{r},\bm{r}')$ to 
functions $F_k(\bm{r}, \bm{r}')$
\begin{eqnarray}
F_{k}(\bm{r},\bm{r}') &=& \sum_{n=1}^{n_k} F(\bm{r}+na_x\hat{x},\bm{r}') e^{-ik(x+na_x-x')}, \label{msfheen}\\
F(\bm{r},\bm{r}') &=& \frac{1}{n_k}\sum_{k} F_k(\bm{r},\bm{r}') e^{ik(x-x')}, \label{msfterug}
\end{eqnarray}
where $n_k$ is the number of equidistant $k$-points in the 1D Brillouin zone (BZ). We use 
periodic boundary conditions: $F(\bm{r}+n_ka_x\hat{x}, \bm{r}') = F(\bm{r}, \bm{r}'+ n_ka_x\hat{x})
=F(\bm{r}, \bm{r}')$. The functions $F_{k}(\bm{r},\bm{r}')$ are fully periodic: 
$F_k(\bm{r}+a_x\hat{x}, \bm{r}') = F_k(\bm{r}, \bm{r}'+ a_x\hat{x})= F_k(\bm{r}, \bm{r}')$,
so that $\bm{r}$ and $\bm{r}'$ can be chosen in the unit cell.
We calculate the one-particle Green function for imaginary times
\begin{equation}
G_k(\bm{r},\bm{r}',i\tau) = \left\{
\begin{array}{rll}\!\! i\!\!&\displaystyle\sum_v u_{vk}(\bm{r})u^*_{vk}(\bm{r}')e^{-(\epsilon_{vk}-\epsilon_F)\tau} &\mbox{ for } \tau < 0,\\
&& \\ 
\!\!-i\!\!&\displaystyle\sum_c u_{ck}(\bm{r})u^*_{ck}(\bm{r}')e^{-(\epsilon_{ck}-\epsilon_F)\tau}&\mbox{ for } \tau > 0, 
\end{array} \right.
\end{equation}
where $u_{nk}(\bm{r}) = \phi_{nk}(\bm{r}) e^{-ikx}$ and $\epsilon_{nk}$ (with $n=c,v$) are 
DFT-LDA wavefunctions and the corresponding energies (with $v$ and $c$
referring to valence and conduction states, respectively).
$\epsilon_F$ is the Fermi energy (set in the middle of the DFT-LDA gap).
 We further calculate the irreducible single-chain polarizability
function in the Random Phase Approximation (RPA)
\begin{equation}
 P_k(\bm{r},\bm{r}',i\tau) = 
- 2i\sum_q G_q(\bm{r},\bm{r}',i\tau)G_{q-k}(\bm{r}',\bm{r},-i\tau),
\label{polariz}
\end{equation}
the screened Coulomb interaction
\begin{equation} W_k(\bm{r},\bm{r}',i\omega) = \left[\tilde{v}^{-1}_k(
\bm{r},\bm{r}')-P_k(\bm{r},\bm{r}',i\omega)\right]^{-1},
\label{wscr}
\end{equation}
where $\tilde{v}_k(\bm{r},\bm{r}')$ is a cut-off Coulomb interaction in mixed 
space, discussed in the next Section,
and we calculate the electronic self-energy
\begin{equation}
\Sigma_k(\bm{r},\bm{r}',i\tau)= i \sum_q G_q(\bm{r},\bm{r}',i\tau)W_{k-q}
(\bm{r},\bm{r}',i\tau).
\label{self-energy}
\end{equation}
We calculate all the above two-point functions
on a double $24\times 24\times 24$ real-space grid for $\bm{r}$ and $\bm{r}'$
in the unit cell. This corresponds
to a plane wave cut-off of 25 Ry.
The total number of valence and conduction bands taken into account was 300.
In Eqs. (\ref{polariz}), (\ref{wscr}) and (\ref{self-energy}) we switch between
time- and frequency
domain using Fourier transforms.
Our imaginary-time grid has an exponential spacing (0.25 a.u.\ near $\tau=0$, up to a spacing
6 a.u.\ near
$\tau_{max}=32.0 {\rm{\mbox{ }a.u.}}$) and we interpolate to a linear grid when using the Fast Fourier Transform (FFT) to
imaginary frequency. A similar exponential grid is used for the imaginary frequency.
We split the self-energy in an exchange part $\Sigma^x$ and a correlation part $\Sigma^c$:
\begin{eqnarray}
\Sigma^x_k(\bm{r}, \bm{r}')&  =& \sum_q iG_q (\bm{r}, \bm{r}', -i\delta) \tilde{v}_{q-k}(\bm{r}', \bm{r}), \label{expart}\\
\Sigma^c_k (\bm{r},\bm{r}',i\tau)& =&  \sum_q iG_q (\bm{r}, \bm{r}', i\tau) W^{\rm{scr}}_{q-k}(\bm{r},\bm{r},i\tau),
\label{corrpart}
\end{eqnarray}
where $\delta$ is an infinitesimally small positive time, and
$W^{\rm{scr}}$ is the screening interaction:
\begin{equation}
W^{\rm{scr}}_k (\bm{r},\bm{r}',i\omega)\equiv W_{k}(\bm{r}, \bm{r}', i\omega)
-\tilde{v}_{k}(\bm{r}, \bm{r}').\label{wscrsplit}
\end{equation}
In the calculation of $\Sigma^x$ we use a 
1D Brillouin-zone sampling of 10 equally spaced $k$-points, 
and in the calculation of $\Sigma^c$ 4 $k$-points (since
the screening interaction is short-ranged, the convergence of $\Sigma^c$
with respect to the number of $k$-points is faster than that of $\Sigma^x$).
With the above parameters, the calculated QP gap has converged to within about 0.05 eV.
From a two-pole fit on the
imaginary-frequency axis an analytical continuation to the real-frequency axis is obtained:
$\Sigma^{c}(i\omega) \rightarrow {\Sigma}^c (\omega)$.\cite{spacetime}
Subsequently, the QP equation Eq.\ (\ref{qpeq}) can be solved by 
replacing the QP wave functions by the DFT-LDA wave functions and obtaining $E_{nk}$
iteratively:
\begin{equation}
E_{nk} = \epsilon_{nk} +\left<\phi_{nk}\left|\Sigma^c_k(E_{nk})+\Sigma^x_k -
V^{xc}\right|\phi_{nk}\right>.
\end{equation}

\subsection{Treatment of the Coulomb interaction}\label{coultreat}

We have developed a novel procedure to deal with
the $\bm{k} = 0$ and $\bm{r} = \bm{r}'$ singularities
of the Coulomb interaction $v_{\mathbf{k}}(\bm{r},\bm{r}')$. 
We will first describe the procedure for a 3D system and later
explain the specific adaptations of this procedure we used 
for our quasi-1D system.

In reciprocal space
the Coulomb interaction is given by:
\begin{equation}
v_{\mathbf{k}}(\bm{K},\bm{K}') = \frac{4\pi}{|{\mathbf{k}}+{\mathbf{K}}|^2}\delta_{\mathbf{K},\mathbf{K}'}.
\label{vk}
\end{equation}
We replace $v_0(0,0)$, which would be infinite in Eq. (\ref{vk}) by a finite value,
which is obtained in the following way. We evaluate the integral over all space 
of the Coulomb interaction multiplied by a Gaussian:
\begin{eqnarray}
I_\alpha &=& \int d^3 \bm{q} \frac{4\pi}{q^2}e^{-\alpha q^2}= 8\pi^2\sqrt{\frac{\pi}{\alpha}},
\end{eqnarray}
and evaluate
the corresponding sum, {\em{excluding}} the singularity for ${\mathbf{K}} = \bm{k} = 0$:
\begin{equation}
S_\alpha = \Delta V \sum_{{\mathbf{k},\mathbf{K}}}^{}\mbox{}\!^\prime
v_{\mathbf{k}}(\bm{K},\bm{K})e^{-\alpha |{\mathbf{k}}+{\mathbf{K}}|^2},
\end{equation}
where $\bm{k} \in$ 1BZ, the first Brillouin zone of the 3D lattice.
$\Delta V$ is the volume per $(\bm{k},\bm{K})$-point 
and the prime indicates that $\bm{k} = \bm{K}=0$ is excluded in this sum.
We now put:
\begin{equation}
v_{{\mathbf{k}}=0 }({\mathbf{K}}=0, {\mathbf{K}}' =0) 
\equiv \lim_{\alpha \rightarrow 0} \left[ I_\alpha - S_\alpha \right].
\end{equation}
Finally, we obtain $v(\bm{r}- \bm{r}')$ by a discrete FFT of $v_{\mathbf{k}}(\bm{K},
\bm{K}')$ to real space.
We find a finite value for $v(\bm{r}-\bm{r}'=0)$,
solving at the same time the problem with the Coulomb singularity for $\bm{r} - \bm{r}' = 0$. 
In the original formulation of the space-time method,\cite{spacetime} the authors
used a grid for $\bm{r}'$ offset with respect the $\bm{r}$-grid in order to avoid this
singularity. 

In order to study a truly isolated chain, which is a quasi-1D system, we have to avoid 
`crosstalk' between periodic images of the chain in the perpendicular directions.
We do this by dividing space into regions of points that are
closer to the atoms of a specific chain than to those of any other.
Subsequently, we cut off the Coulomb interaction $v(\bm{r}-\bm{r}')$, obtained
in the way described above, by setting it zero if 
$\bm{r}$ and $\bm{r}'$ belong to different regions. Thus we obtain an 
interaction $\tilde{v}(\bm{r}, \bm{r}')$.
In the construction of the Coulomb interaction $v(\bm{r}-\bm{r}')$, 
we take a regular grid
of $\bm{k}$-points with a spacing in the $y$- and $z$-direction approximately equal to that in
the $x$-direction.
From the cut-off interaction $\tilde{v}(\bm{r},\bm{r}')$
we obtain $\tilde{v}_k(\bm{r},\bm{r}')$ in mixed space
from Eq. (\ref{msfheen}) with $k$ now in the 1D Brillouin zone.

\subsection{{The Bethe-Salpeter equation}}
The two-body electron-hole Schr\"odinger equation related to the BSE is solved by expanding the 
exciton wave functions $\Phi(\bm{r}_e, \bm{r}_h)$ 
in products of conduction $\phi_{ck}(\bm{r}_e)$
and valence wave functions $\phi_{vk}(\bm{r}_h)$:\cite{ben,ar,rl,shamrice,strinati}
\begin{equation}
\Phi(\bm{r}_e, \bm{r}_h) = \sum_{k,c,v} A_{kcv}
 \phi_{ck}(\bm{r}_e)\phi^{*}_{vk}(\bm{r}_h).
\end{equation}
Here we have restricted our discussion to excitons which have zero total momentum,
since only these are optically active.
As we are interested in the lowest lying excitons,
an expansion in  the highest occupied valence ($\pi$) and lowest unoccupied conduction ($\pi^{*}$)
bands is sufficient to converge the exciton energies to within 0.1 eV; energy
differences are converged even better. Below, we will give all energies in eV
with a precision of two decimal places.
The exciton binding energies $E_b$ follow from the Schr\"odinger-like equation:\cite{ben,ar,rl}
\begin{equation}
[E_{ck} - E_{vk} - E_g + E_b ] A_{kcv} + \sum_{k'c'v'} 
[2V^x_{kcv,k'c'v'}\delta_{s,0}-W_{kcv,k'c'v'}]A_{{k'c'v'}} = 0,
\label{excitons}
\end{equation}
where $E_g$ is the QP band gap, $E_b$ the exciton binding energy and
$W_{kcv,k'c'v'}$ are the matrix elements of the static ($\omega = 0$)
screened interaction
\begin{equation}
W_{kcv,k'c'v'}=\int\! d^{3}\bm{r}\int\!
d^{3}\bm{r}' \phi _{vk}(\bm{r}')\phi _{ck}^{*}
(\bm{r})W(\bm{r},\bm{r}', \omega =0)\phi _{v'k'}^{*}(\bm{r}')\phi_{c'k'}(\bm{r}), \label{matelem1}
\end{equation}
and $V^x_{kcv,k'c'v'}$ the exchange matrix elements (present for singlet excitons, $s=0$, only)
of the bare Coulomb interaction:
\begin{equation}
V_{kcv,k'c'v'}^{x}= \int\! d^3\bm{r}\int\! d^3\bm{r}'\phi _{v'k'}^{*}(\bm{r})
\phi _{ck}^{*}(\bm{r}')
\tilde{v}(\bm{r},\bm{r}') \phi _{vk}
(\bm{r}')\phi _{c'k'}(\bm{r}).   \label{matelem2}
\end{equation}
The integrals over space in Eqs. (\ref{matelem1}) and (\ref{matelem2}) are
in the calculations replaced by summations over our real-space grid.
We use wave functions
and energies on a grid of 8 $k$-points and extrapolate to a grid of 100 $k$-points.

Formally, dynamical
screening effects may only be ignored in the BSE if $E_g \gg E_{b}$. However,
since it has been shown that
dynamical effects in the electron-hole screening and in the
one-particle Green function largely cancel each other,\cite{bechstedt} this approximation
is nevertheless valid, even if the relation $E_g \gg E_{b}$ does not strictly hold.

We calculate an approximate 
exciton size $a_{\rm{ex}}$ by fitting the exciton coefficients $A_{kcv}$
to the hydrogen-like form:
\begin{equation}
A_{kcv} = \frac{A_{k=0,cv}}{(1+a^2_{\rm{ex}}k^2)^2}. \label{excsize}
\end{equation}
Note that in fact the exciton is highly anisotropic. Nevertheless, Eq.\ (\ref{excsize})
gives pretty good fits and can be used to get a qualitative impression
of the (relative) size of the excitons.

\subsection{Inclusion of interchain screening}\label{diel}

As mentioned in the introduction, in a quasi-1D system, such as an isolated chain
of a polymer in vacuum, there is no long-range screening. For a meaningful comparison
of our calculations to the experimental data, which are obtained from either films or bulk
polymer material, both the intra- and the interchain screening are important, and
only the latter is long-ranged.
It would be desirable to perform a $GW$A and exciton calculation for a 3D crystal structure 
of PT,
but the amount of computational work involved is as yet prohibitively large. 
Since PT samples prepared in many different ways show very similar optical behavior,
we expect the details of the interchain screening not to be extremely important. 
This consideration leads us to propose  
the following approximation for the interchain screening 
interaction, defined analogously to Eq. (\ref{wscrsplit}):
\begin{equation}
W^{\rm scr}_{\rm{inter}}(\bm{r},i\omega) = (1-e^{-r/r_{\rm{inter}}})
\left\{ \left[
         \varepsilon_\bot^2(i\omega)x^2+
\varepsilon_{||}(i\omega)\varepsilon_\bot(i\omega)(y^2+z^2)\right]^{-1/2} -v(\bm{r}) \right\},
\label{weps}
\end{equation}
where $\varepsilon_\bot(i\omega)$ and $\varepsilon_{||}(i\omega)$ are the {\em{ab-initio}}
frequency-dependent
dielectric constants perpendicular and parallel to the chain, respectively.
The counter-intuitive combination of dielectric constants and 
coordinates in Eq. (\ref{weps}) results from solving the Laplace
equation for a point charge in a homogeneous, anisotropic
medium with dielectric constants $\varepsilon_{||}$ and $\varepsilon_{\bot}$.\cite{ll}
The prefactor takes care of a smooth cut-off
for distances smaller than $r_{\rm{inter}}$, for which the interchain screening should be
replaced by the intrachain screening. 
Eq.\ (\ref{weps}) has the correct behaviour for distances larger than the interchain distance
$r_{\rm{inter}}$, for which we take 10 a.u., which is typical for the experimental
crystal structures
of Refs.\ \onlinecite{xtal} and \onlinecite{mo}. 

The total screened interaction for the bulk system then becomes:
\begin{equation}
W_{\rm{total}}(\bm{r}, \bm{r}', i \omega) = W^{\rm{scr}}_{\rm{intra}}(\bm{r}, \bm{r}', i \omega) +
W^{\rm{scr}}_{\rm{inter}}(\bm{r}-\bm{r}', i\omega)
+ v(\bm{r}- \bm{r}'), \label{totalw}
\end{equation}
where $W^{\rm{scr}}_{\rm{intra}}$ is the intrachain screening already calculated 
with Eq.\ (\ref{wscr}).
The screened interaction 
$W_{\rm{total}}$ is
correct at short range, where the interchain screening is vanishingly small compared
to the $1/r$-divergence of the intrachain screening,
and at long range, where the intrachain screening vanishes due to its quasi-1D nature.
Of course, for intermediate ranges, it is not strictly allowed to simply add the
parts representing long- and short-ranged screening, but
we expect Eq.\ (\ref{weps}) to give a reasonable interpolation there.
 
Note that the interchain screening part given by Eq.\ (\ref{weps}) is  long-ranged by construction,
and 8 $k$-points are now needed to converge the
corresponding self-energy $\Sigma^c_{\rm{inter}}$ from Eq. (\ref{corrpart}).
On the other hand, the required number of
real-space grid points in order to calculate $\Sigma^c_{\rm{inter}}$ is less than before,
because $W_{\rm{inter}}^{\rm{scr}}$
is a very smooth function of $\bm{r}$; a $12\times 12\times 12$ real space grid turns out
to be sufficient.
The total self-energy can be expressed as:
\begin{equation}
\Sigma_{\rm{total}} = \Sigma^c_{\rm{intra}} + \Sigma^c_{\rm{inter}} + \Sigma^x.
\end{equation} 
Because the self-energies in this equation are additive, we can reuse the
self-energies $\Sigma^c_{\rm{intra}}$ and $\Sigma^x$, which we have already calculated for the
isolated chain.

The overlap between wave functions, and therefore the {\em{electronic}} 
coupling between
neighboring chains, is very small.
This means that we can use the isolated-chain wave functions to calculate
the Green function and self-energy. This obviously implies
that in our exciton calculations we restrict ourselves to excitons in which 
we take the electron and hole are on the same chain (so-called intrachain excitons). 
In summary, the only, but essential, difference between our calculations for the isolated
PT chain and bulk PT is in the use of an interchain screened interaction.

\subsection{Dielectric tensor of crystalline PT}

In order to construct the screened interaction of Eq.\ (\ref{weps}),
we have to calculate the dielectric tensor of bulk PT.
We do this for the crystalline structure of Ref. \onlinecite{xtal},
which is reproduced in Fig.\ \ref{xtal2}. We use a model in which the chains
are replaced by polarizable line objects with a polarizability tensor
obtained from the single-chain polarizability function.
The principal axes of the chain are the following:
\begin{equation}
\hat{x}_1 = \hat{x}, \mbox{  } \hat{x}_2 = \frac{1}{\sqrt{2}}(\hat{z}+\hat{y}),\mbox{  }
\hat{x}_3 = \frac{1}{\sqrt{2}}(\hat{z}-\hat{y}).
\end{equation}
The full polarizability function $X(\bm{r},\bm{r}', i\omega)$ of a single chain
is given by:
\begin{eqnarray}
X(\bm{r},\bm{r}', i\omega)  &=&
 P(\bm{r},\bm{r}', i\omega) +
 \int\!\!\!\int d\bm{r}''d\bm{r}''' P(\bm{r},\bm{r}'', i\omega) 
W_{\rm{intra}}(\bm{r}'',\bm{r}''',
i\omega) P(\bm{r}''',\bm{r}', i\omega) \\
&\equiv& X^{(0)}(\bm{r},\bm{r}', i\omega)  +X^{(1)}(\bm{r},\bm{r}', i\omega) .
\end{eqnarray}
The long-wavelength ($q\rightarrow 0$) polarizability tensor ${\underline{\chi}}$ 
per unit of chain length
of a single chain in the $(x_1,x_2,x_3)$ coordinate system is diagonal
and has diagonal elements given by:
\begin{equation}
\chi_1(i\omega) = \lim_{q\rightarrow 0} \left[ \frac{1}{q^2}\int\!\!\!\int d\bm{r}d\bm{r}'
e^{-iq(x_1-x_1')}X(\bm{r}, \bm{r}',i\omega) \right], \label{chi1}
\end{equation}
and for $j$=2,3:
\begin{equation}
\chi_j(i\omega) = \int\!\!\!\int d\bm{r}d\bm{r}'
x_jX(\bm{r}, \bm{r}',i\omega)x_j'. \label{chirest}
\end{equation}
The calculation of $\chi_j(i\omega)$ has been performed with 4 $k$-points, with the exception
of $\chi_1^{(0)}$, for which it proved to be necessary to use 8 $k$-points.

If we now approximate the chains by polarizable line objects with the above
polarizability tensor, we can calculate 
the macroscopic dielectric tensor of the crystal\cite{xtal} of these chains.
This is done by a procedure of which the details are given in
the Appendix.
The axes of the crystal unit cell are denoted by $\hat{a}$, $\hat{b}$ and $\hat{c}=\hat{x}
=\hat{x}_1$.
Dropping the frequency dependence in the notation, we find the following
expression for $\varepsilon_{c}$:
\begin{equation}
\varepsilon_{c} = 1+\frac{4\pi\chi_{1}}{A}, \label{epsx}
\end{equation}
where $A$ is the surface area per chain in the plane perpendicular to the chain.
For $\varepsilon_{a}$ and $\varepsilon_{b}$ we find:
\begin{equation}
\varepsilon_{\gamma} = \frac{1}{1 - {\frac{4\pi}{A} \tilde{\chi}_\gamma}},
\label{epsall}
\end{equation}
where $\gamma=a,b$ and $\tilde{\chi}_\gamma$ is the effective polarizability
of the chain along the
$\gamma$-axis.
In the Appendix details of the calculation of $\varepsilon_a$, $\varepsilon_b$, and
$\varepsilon_c$ are given.

To retain the tetragonal symmetry in our calculation (in order keep the computations
feasible),
we average $\varepsilon_{a}(i\omega)$ and $\varepsilon_{b}(i\omega)$, which are not
very different, to obtain 
$\varepsilon_{\bot}(i\omega)$. For $\varepsilon_{||}(i\omega)$ we take $\varepsilon_{c}(i\omega)$.
Note that for using the screened interaction of Eq. (\ref{weps}) to in the implementation
of the $GW$A formalism presented in Section~\ref{GWimpl}, we have calculated the dielectric constants
$\varepsilon_{\bot}(i\omega)$ and $\varepsilon_{||}(i\omega)$ 
along the imaginary frequency axis.

\section{{Results}}\label{Results}
\subsection{Isolated chain}
\begin{figure}
\leavevmode
\centerline{
\epsfysize10cm
\epsffile{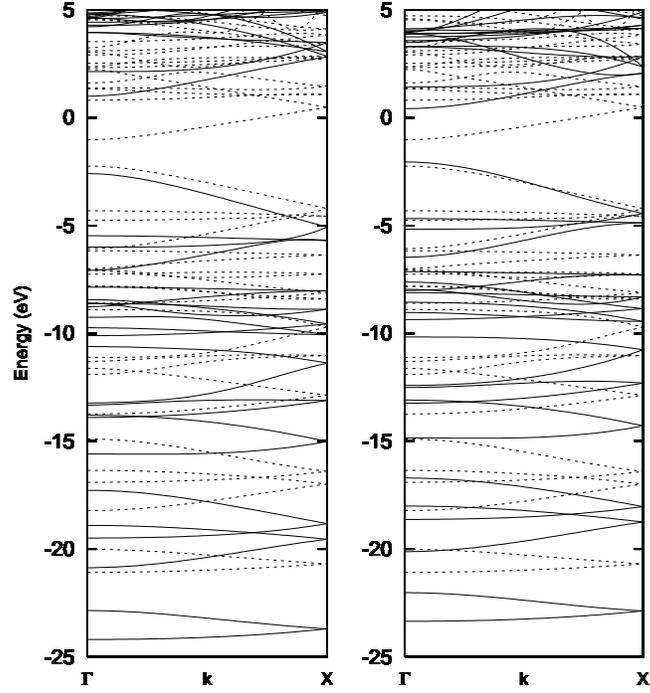}}
\caption{The quasi-particle band structure of both
an isolated chain of PT (full lines, left) and  bulk PT (full lines,
right) compared to the DFT-LDA band structure (dashed lines in both pictures).}
\label{qpboth}
\end{figure}

The calculated $GW$A QP band structure (together with the DFT-LDA band structure)
is shown in Fig.~\ref{qpboth}, left panel.  We find a minimal band gap $E_g$ at $\Gamma$
of 3.59 eV, which is quite large compared to the DFT-LDA value of 1.22 eV. 
The effective masses, $m^* = 1/\hbar^2(\partial^2E/\partial k^2)^{-1}$,
of the $\pi$ and $\pi^*$ bands at $\Gamma$, which
are 0.15 and 0.17 $m_e$ (with $m_e$ the free electron mass)
in DFT-LDA, are reduced by about 15\% in the $GW$A to
0.13 and 0.15  $m_e$. This corresponds to an increase of the band width
from 1.96 and 1.51 eV in DFT-LDA to 2.48 and 1.81 eV in the $GW$A, for the
$\pi$ and $\pi^{*}$ bands, respectively. In an earlier
$GW$A study, a similar increase of the bandwidth was found for a wide variety 
of materials.~\cite{bwsh}

\begin{figure}
\leavevmode
\centerline{
\epsfysize10cm
\epsffile{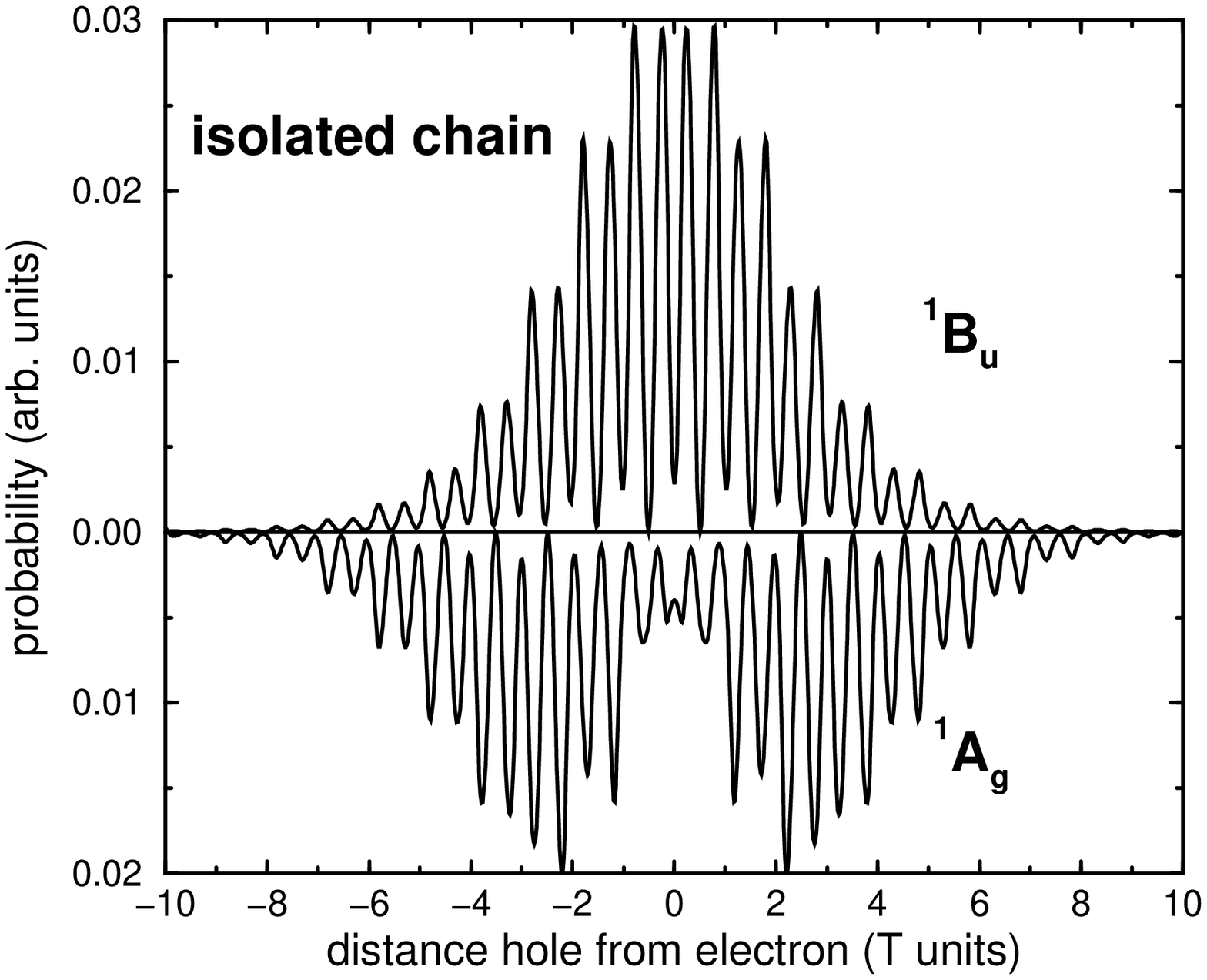}}
\centerline{\epsfysize10cm
\epsffile{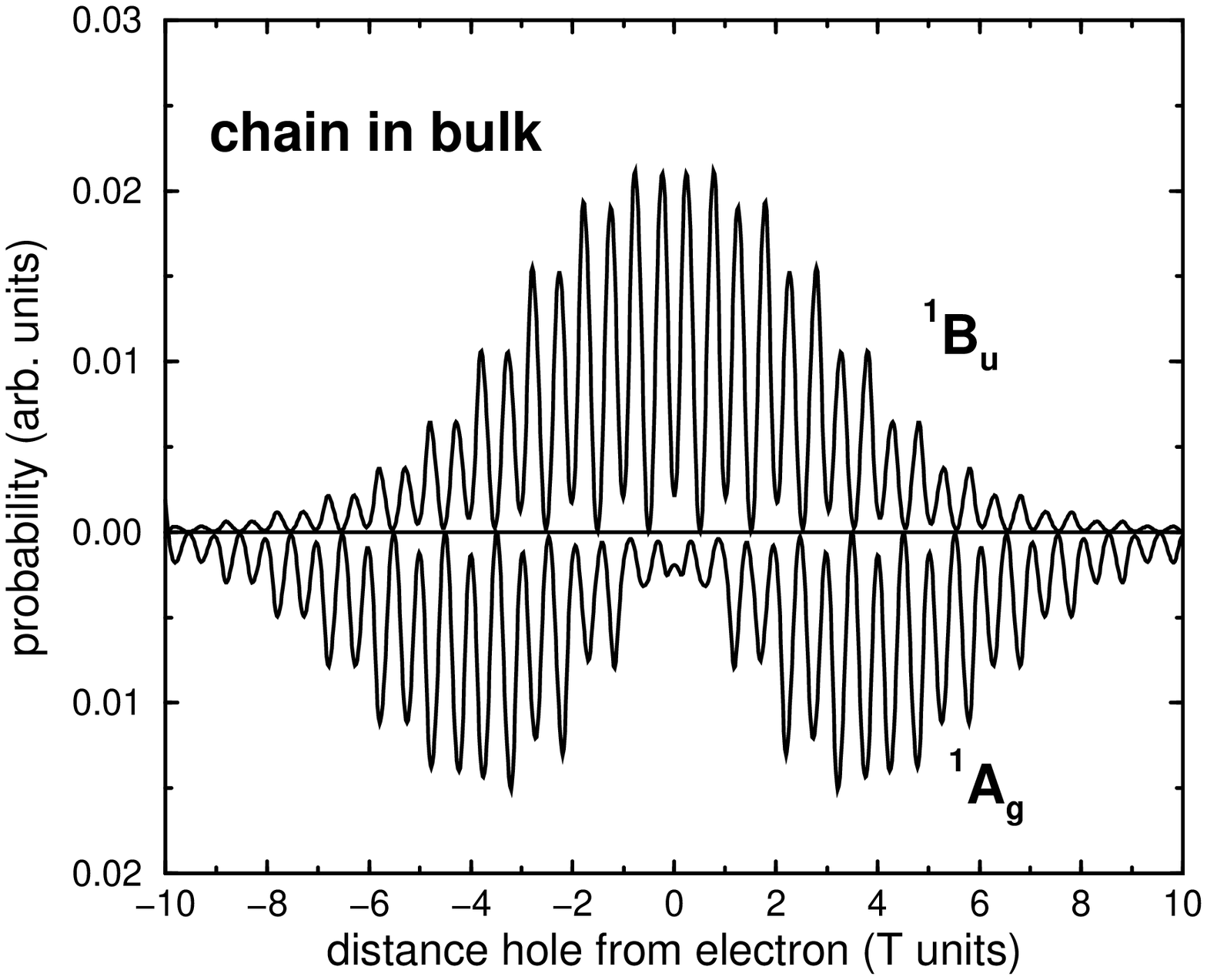}}
\caption{The hole probability along the chain with the electron fixed at 1 a.u.\ above
the inversion center for the two lowest singlet excitons, $^1$B$_u$ (above axis) and $^1$A$_g$
(below axis), for both the isolated chain (top) and the bulk situation (bottom).}
\label{excwave}
\end{figure}

The lowest-lying singlet exciton ($^1$B$_{\rm{u}}$)
has a binding energy $E_b$ of 1.85 eV. The size $a_{\rm{ex}}$
of this exciton, calculated using Eq.\ (\ref{excsize}),
is 12 a.u., i.e.\ less than two thiophene rings.
To give an impression of the exciton wave
function $\Phi(\bm{r}_e, \bm{r}_h)$, we have plotted in Fig.\ \ref{excwave} (top panel)
the probability to find the hole at a distance $x_h$ along the chain from 
the electron,
\begin{equation}
{\rm{Prob}} (x_h) \sim  \int \!\! dy_hdz_h |\Phi(\bm{r}_e, \bm{r}_h) |^2,
\end{equation}
where the electron coordinate $\bm{r}_e$ is taken 1 a.u.\ from the inversion center,
in the direction perpendicular to the polymer plane (for the electron coordinate
{\em{in}} the inversion center, this probability would be zero due to symmetry). 
We have plotted ${\rm{Prob}} (x_h)$ for both the $^1$B$_{\rm{u}}$ and $^1$A$_{\rm{g}}$
excitons.

As the optical gap is given by $E_o = E_g - E_b$, 
we have $E_o = 1.74$~eV,
in good agreement with
the experimental value of 1.8 eV~\cite{sakurai} (see Table~\ref{excagain}). 
While there is good agreement for the optical gap,
the {\em difference} between the $^1{\rm{B}}_{\rm{u}}$ and $^1{\rm{A}}_{\rm{g}}$
binding energies of the isolated PT chain is definitely
{\em{not}} in agreement with experiment,~\cite{sakurai} see Table~\ref{excagain}. 
Moreover, the $^1{\rm{B}}_{\rm{u}}$ exciton binding energy of 1.85 eV is very
large compared to values currently discussed in the literature, which range
from $\sim 0.1$ to $\sim 1.0$ eV.~\cite{bredasheeger}

\subsection{{Dielectric properties}}
We calculate the polarizabilities per unit length
$\chi_{j}(i\omega)$ with Eqs.~(\ref{chi1}) and (\ref{chirest}).
The obtained $\omega = 0$ values
are listed in Table~\ref{poleps}. Note that the polarizability along the chain, i.e.\ in
the direction of the extended carbon $\pi$-system, is much larger than those in the
perpendicular directions. This difference is reflected in the dielectric constants
$\varepsilon_{\gamma}(i\omega)$ 
calculated using Eqs.~(\ref{epsx}) and (\ref{epsall}); 
the dielectric constant along the chain is much larger than those in the perpendicular
directions. 
In real systems with disorder the conjugation length will be finite, which will 
reduce $\varepsilon_{||}$. Note, however, that the {\it{perpendicular}} dielectric constant
$\varepsilon_{\bot}$ plays the dominant role in the interchain
screening of Eq.\ (\ref{weps}) {\em{along}} the chain.

\subsection{Crystalline polythiophene}

The resulting band structure, calculated using the bulk screening from Eqs.\ (\ref{weps}) and
(\ref{totalw}) is given in Fig.\ \ref{qpboth}, right panel. The QP gap $E_g$ has decreased to
2.49 eV; the $^1$B$_{\rm{u}}$ exciton binding energy is 0.76 eV (see Table \ref{excagain}).
 Hence, the predicted optical gap is 1.73 eV, virtually unchanged from the isolated
chain results of 1.74 eV and in good agreement with experiment.~\cite{sakurai}
Note that the {\em{absorption}} gap
of Ref.\ \onlinecite{sakurai} is 2.0 eV, also found in earlier work on PT,~\cite{kobi} but
the {\em{luminescence}} gap is 1.8 eV.
There are two reasons why we should compare our result to the latter gap.
The first reason is that absorption occurs everywhere in a sample, both in the ordered 
and disordered parts, but luminescence occurs predominantly in the most ordered parts with
the longest conjugation lengths. This is because, prior to recombination, excitons 
diffuse to those parts of the sample where they have the lowest energy.\cite{absvslum}
The second reason is that after photoexcitation, the rings, which may be twisted around their
common C-C bond, tend to co-planarize in the excited state, due to the fact that the excited 
state is slightly more quinoid than the aromatic ground state.~\cite{bredasheeger}
As we are performing our calculations for a perfect, co-planar chain of PT,
we should therefore compare our optical gap to the luminescence gap. 
Note that in principle it is possible that excitons trapped in defects or disordered parts of
the sample to have a lower energy than in a fully conjugated, defect-free polymer. 
However, the luminescence spectrum of Ref.\ \onlinecite{sakurai} can be fully understood in 
terms of the $^1$B$_{\rm{u}}$ exciton 
decay and its vibronic side bands, which means that such defects are either rare 
or that excitons trapped by such defects decay non-radiatively.

What is very important, is that the
{\em{relative}} exciton energies (also listed in Table \ref{excagain}) are now also in
good agreement with experiment. The sizes of the excitons
have increased by $\sim 50$\%; the $^1$B$_{\rm{u}}$ size $a_{\rm{ex}}$ is
now 18 a.u., or slightly more than two rings. In Fig. \ref{excwave} (bottom panel)
it is clearly seen that the excitons are larger than the corresponding excitons
on the isolated chain (top panel).

In order to test the sensitivity of our results to the precise value of the cutoff
distance $r_{\rm{inter}}$ in Eq.~(\ref{weps})
we performed similar calculations for $r_{\rm{inter}}$~=~8~a.u.\
and $r_{\rm{inter}}$~=~12~a.u. These data are also listed in Table~\ref{excagain}.
The QP band gaps are 2.32 and 2.61~eV, respectively. 
The $^1{\rm{B}}_{\rm{u}}$ binding energies are 0.64 and 0.86~eV and hence the
optical gaps are 1.68 and 1.73~eV, respectively. This means that the optical gap
is quite insensitive to the choice of $r_{\rm{inter}}$.
This is consistent with
the fact that in the limit $r_{\rm{inter}}\rightarrow \infty$, which 
corresponds to no interchain screening, we should find the isolated chain
absorption gap of 1.74 eV. The energy differences between
the excitons are even less sensitive to $r_{\rm{inter}}$.
The good agreement with experiment and the fact that especially the
optical gap and the energy separation between the excitons
are hardly influenced by varying $r_{\rm{inter}}$ are also {\em{a posteriori}} 
justifications for our model screening interaction Eq. (\ref{totalw}).

\section{{Conclusions and discussion}}\label{Discuss}

Summing up, we have calculated the 
quasi-particle band structure and lowest-lying exciton binding energies of an isolated
polythiophene chain and crystalline polythiophene.
For the isolated chain (where there is only intrachain screening) we find a large band gap and
large exciton binding energies, due to the absence of long-range screening. 
After including interchain screening, which is responsible for the long-range screening
in bulk polythiophene, we find that both the band gap and exciton binding energies 
are drastically reduced. However, the optical gap is hardly affected.
We suggest that these conclusions hold for conjugated polymers in general.

This sheds light on the fact that the calculations by Rohlfing and Louie~\cite{oeicp}
on isolated chains of PA and PPV yield good results for the optical gaps, whereas their
lowest-lying singlet exciton binding energy of 0.9 eV for PPV is in excess of recent 
experimental values of $0.35\pm 0.15$~eV,~\cite{intermediate} obtained by a direct
STM measurement for an alkoxy-substituted PPV, and $0.48 \pm 0.14$ eV for unsubstituted
PPV.\cite{alvpriv}
The inclusion of interchain screening effects will drastically reduce
their calculated binding energy and may
well lead to agreement with this experiment. 
Clearly, it would also be very interesting
to repeat the experiment in Ref.~\onlinecite{intermediate} for polythiophene and polyacetylene.
Interestingly enough, a value of 0.4 eV is obtained for the exciton binding energy in PPV by means
of an  effective-mass appromixation in which the electron-hole interaction is
derived from a bulk dielectric tensor.\cite{gdc2}
The difference of about a factor of two in exciton binding energy between crystalline
PT and PPV can, at least qualitatively,
be explained by the differences in reduced masses $\mu$
($1/\mu = 1/m_{\pi}+1/m_{\pi^*}$)
of PT and PPV, for which we find $\mu^{\rm{PT}} = 0.08 m_e$, while
$\mu^{\rm{PPV}} =0.04 m_e$,~\cite{gdc2} both
in DFT-LDA, and by the fact that in an effective-mass approximation
the binding energy is proportional to $\mu$. Of course, these arguments, which
are qualitative only, do not take away the need for {\em{ab-initio}}
calculations on the crystalline phase of PPV.

Further, the apparent discrepancy of the results for PA by Ethridge {\em{et al.}}\cite{ethridge}
and those of Rohlfing and Louie,~\cite{oeicp} can be understood.
The latter find, for an isolated chain, a QP gap of 2.1 eV and an exciton
binding energy of 0.4 eV, yielding an absorption gap of 1.7 eV.
The former find a QP gap of 1.86 eV and do not include excitonic effects. 
This calculation, however, is performed for
one PA chain in the same volume as a PA chain in a crystal would have.
Therefore,
this calculation is in fact one for a bulk situation, which means that this
QP gap is by our arguments expected to be smaller than that of Rohlfing and Louie. 
Furthermore, our arguments predict an exciton binding energy in bulk PA considerably 
smaller than the 0.4 eV of Rohlfing and Louie. 

We conclude that a correct many-body description of the electronic and optical
properties of bulk polymer systems should include the effect of interchain screening. 
An important consequence of this conclusion is that neither Hartree-Fock nor DFT-LDA calculations 
should be relied upon in this context, since Hartree-Fock does not contain
screening effects at all and since the exchange-correlation
potential in DFT-LDA only depends on the local density and cannot 
describe the non-local effects due to the long-range screening. Moreover, 
since exciton effects play such a large role in conjugated polymers, it is
essential to take these effects into account.

\section*{Acknowledgements}
Financial support from NCF (Nationale Computer Faciliteiten) project SC-496 is acknowledged. 
G.B. acknowledges the financial support from Philips Research
through the FOM-LZM (Fundamenteel Onderzoek der Materie - Laboratorium Zonder Muren) program.

\appendix\section{Calculation of the crystal dielectric tensor within a line-dipole model}
We apply an electric field $\bm{E}_{\rm{appl}}
(\bm{r}) = \bm{E}_0 e^{i\mathbf{k}\cdot\mathbf{r}}$ (and we will take the limit $k \rightarrow 0$),
where $\bm{E}_0$ and $\bm{k}$ are parallel to the $a$,$b$ or $c$-axis of the crystal (see Fig.~\ref{xtal2})
to calculate $\varepsilon_a$, $\varepsilon_b$ and $\varepsilon_c$,
respectively. 
The applied field $\bm{E}_{\rm{appl}}$ leads to an induced field
$\bm{E}_{\rm{ind}}(\bm{r})$; the total microscopic
field $\bm{E}_{\rm{micr}}(\bm{r})$ is then given
by:
\begin{equation}
\bm{E}_{\rm{micr}}(\bm{r}) = \bm{E}_{\rm{appl}}(\bm{r}) +\bm{E}_{\rm{ind}}(\bm{r}).
\end{equation} 
We define $\vec{\rho} = u\hat{a}+v\hat{b}$ with
$\rho^2 = u^2 + v^2$. Note that there are two different chains: the ${\mathcal{A}}$ type,
at the corners of the unit cell, and the ${\mathcal{B}}$ type at the center of the unit cell.
For the ${\mathcal{A}}$ and ${\mathcal{B}}$ chain we have:
\begin{eqnarray}
\bm{p}_{\mathcal{A}}(x) &=& {\underline{\chi}}_{\mathcal{A}} \cdot \bm{E}'_{\rm{micr}}
(x,\vec{\rho}=0) \label{pA}\\
\bm{p}_{\mathcal{B}}(x) &=& {\underline{\chi}}_{\mathcal{B}} \cdot \bm{E}'_{\rm{micr}}(x,\vec{\rho}=
\frac{1}{2}\hat{a}+\frac{1}{2}\hat{b})
\label{pB}
\end{eqnarray}
with $\bm{p}_{\mathcal{A}}(x)$ ($\bm{p}_{\mathcal{B}}(x)$) 
the long-wavelength dipole moment
per unit length of the $\mathcal{A}$ ($\mathcal{B}$)
chain and ${\underline{\chi}}_{\mathcal{A}}$ (${\underline{\chi}}_{\mathcal{B}}$)
the polarizability tensor of the ${\mathcal{A}}$ (${\mathcal{B}}$)
chain calculated with Eqs.\ (\ref{chi1}) and (\ref{chirest}) and using the relations:
\begin{equation}
{\underline{\chi}}_{\mathcal{A}} = {\underline{U}}^{-1}_{\mathcal{A}} \cdot
{\underline{\chi}}\cdot{\underline{U}}_{\mathcal{A}},\mbox{  }
{\underline{\chi}}_{\mathcal{B}}=
{\underline{U}}^{-1}_{\mathcal{B}}\cdot{\underline{\chi}}\cdot{\underline{U}}_{\mathcal{B}},\label{chiB}
\end{equation}
with ${\underline{U}}_{\mathcal{A}}$ and ${\underline{U}}_{\mathcal{B}}$ are the
rotation matrices relating the ($x_2$,$x_3$) coordinate system to the
($a$,$b$) coordinate system ($\hat{c}=\hat{x}_1$):
\begin{eqnarray}
{\underline{U}}_{\mathcal{A}} & =&
 \left( \begin{array}{rr} \cos(\frac{\pi}{4}-\alpha) & -\sin(\frac{\pi}{4}-\alpha)\\
\sin(\frac{\pi}{4}-\alpha)& \cos(\frac{\pi}{4}-\alpha) \end{array} \right) ,\\
{\underline{U}}_{\mathcal{B}} & =& \left( \begin{array}{rr} \cos(\frac{3\pi}{4}-\alpha) & \sin(\frac{3\pi}{4}-\alpha)\\
-\sin(\frac{3\pi}{4}-\alpha)& \cos(\frac{3\pi}{4}-\alpha) \end{array} \right).
\end{eqnarray}
The prime in Eqs.\ (\ref{pA}) and (\ref{pB}) indicates that the field caused
by the chain itself is excluded.
We will refer to our model, in which a PT chain is represented by an homogeneous line with a certain dipole moment per unit
length $\bm{p}$, as a `line-dipole'.

In CGS units the dielectric tensor ${\underline{\varepsilon}}$ is defined as:
\begin{equation}
\bm{E}(\bm{r})+4\pi\bm{P}(\bm{r}) = {\underline{\varepsilon}}\cdot \bm{E}(\bm{r}).\label{eqforeps}
\end{equation}
where $\bm{E}(\bm{r})$ is the macroscopic field, and ${\bm{P}}(\bm{r})$ is the macroscopic 
polarization. For each direction of the applied field, we will calculate
$\bm{E}_{\rm{ind}}(\bm{r})$, evaluate the macroscopic fields $\bm{E}(\bm{r})$ and
$\bm{P}(\bm{r})$ by averaging, and solve Eq.\ (\ref{eqforeps}) to
obtain the dielectric tensor $\underline{\varepsilon}$. 

\subsection{Calculation of $\varepsilon_c$}

For $\bm{E}_{\rm{appl}}$ and $\bm{k}$ parallel to $\hat{x}$ (and hence to $\hat{c}$ and also $\hat{x}_1$), we have
for both the ${\mathcal{A}}$ and ${\mathcal{B}}$ chain from Eqs. (\ref{pA}) and (\ref{pB}):
\begin{equation}
p_x(x) = \chi_{1} E'_x(x) \label{pxeqchiE}
\end{equation}
The field induced by a line-dipole on the $x$-axis is given by:
\begin{equation}
\bm{E}_{\rm{ind}}(\bm{r}) = -\vec{\nabla} \Phi (\bm{r}) = - {\vec{\nabla}}\int\ 
\frac{p_xe^{ikx'}(x-x')}{|\bm{r}-\bm{r}'|^3} dx',\label{eindintx}
\end{equation}
where $\Phi$ is the electrostatic potential and we have used the fact that $p_x(x') = p_xe^{ikx'}$.
Evaluation of Eq.\ (\ref{eindintx}) yields:
\begin{equation}
E_{{\rm{ind}} ,x}(\bm{r}) =-2k^{2}p_{x}K_{0}(\rho k)e^{ikx},  \label{Eindx} \\
\end{equation}
where $K_0$ is a zeroth order Bessel function of the third kind.
From here on, we omit the factor $e^{ikx}$.
We can calculate the total microscopic field at the $x$-axis, due to both applied and induced fields, 
for a crystal of line-dipoles, by summing over all line-dipoles but the one at the origin:
\begin{equation}
E'_{{\rm{micr}},x} (\vec{\rho}=0)= E_{{\rm{appl}},x}(\vec{\rho}=0) + \sum_{\vec{\rho_i}\neq 0} E_{{\rm{ind}} ,x}(-{{\vec{\rho}}}_i) 
\label{exsum}
\end{equation}
where the positions of the other chains are given by $\vec{\rho_i}$.
In the limit $k \rightarrow 0$, we can replace the sum by an integral:
\begin{eqnarray}
\lim_{k\rightarrow 0} \sum_{\vec{\rho_i}\neq 0} E_{{\rm{ind}} ,x}(\vec{\rho_i}) &=&
 \frac{2\pi p_x}{A} \int_0^\infty \rho' d\rho' K_0 (\rho')\\
& =& -\frac{4\pi}{A}p_x, \label{zomaareen}
\end{eqnarray}
where $\rho'$ = $\rho k$ and $A = ab/2$ is the area of the two dimensional unit cell per chain.
Substitution of Eq.\ (\ref{zomaareen}) and (\ref{exsum}) in Eq. (\ref{pxeqchiE}) yields:
\begin{equation}
p_{x}=\frac{\chi _{1}A}{A+4\pi \chi _{1}}E_{\rm{appl},x}.  \label{dipmomxopgelost}
\end{equation}
Since $P_x=p_{x}/A$, we have:
\begin{equation}
P_x=\frac{\chi _{1}}{A+4\pi \chi _{1}}E_{{\rm{appl}},x}.  \label{Popgelostx}
\end{equation}
The macroscopic field $E_x$ is the average over the two-dimensional unit cell of the 
microscopic field as given by Eq.\ (\ref{exsum}) for general $\vec{\rho}$, but now including
the chain at $\vec{\rho_i} = 0$ in the sum:
\begin{eqnarray}
E_{x}&=&E_{{\rm{appl}},x}+\lim_{k \rightarrow 0}\frac{1}{2A}\int_{{\rm{unit\mbox{ }cell}}} d^{2}\vec{\rho} 
\sum_{\vec{\rho_{i}}} E_{{\rm{ind}},x}(\vec{\rho}-\vec{\rho_i}) \\
    &=&E_{{\rm{appl}},x}+\lim_{k \rightarrow 0}\frac{2\pi}{A}\int \rho d\rho
\sum_{\vec{\rho}_{i}} E_{{\rm{ind}},x}(\vec{\rho}-\vec{\rho_i}) \\
    &=& E_{{\rm{appl}},x}-\frac{4\pi p_x}{A}
\end{eqnarray}
Combining this with Eqs.\ (\ref{eqforeps}) and (\ref{Popgelostx})
we obtain Eq.\ (\ref{epsx}):
\begin{equation}
\varepsilon _{c}=1+ \frac{4\pi\chi _{1}}{A}.
\end{equation}

\subsection{Calculation of $\varepsilon_a$ and $\varepsilon_b$}

We now take $\bm{E}_{{\rm{appl}}}(\bm{r})$ and $\bm{k}$ parallel to $\hat{a}$. The
derivation for $\bm{E}_{{\rm{appl}}}(\bm{r})$ and $\bm{k}$ parallel to $\hat{b}$ is
equivalent. 
The dipole moments of the chains must satisfy Eqs.\ (\ref{pA}) and (\ref{pB}).
The field induced by the chain at the origin is given by:
\begin{eqnarray}
\bm{E}_{\rm{ind}}(\bm{r}) = - \nabla \Phi(\bm{r})
& =& -\nabla \int \frac{\bm{p}_{{\mathcal{A}}}\cdot \bm{r} }{|\bm{r}
-\bm{r}'|^3}dx' \\
                      & =& {\underline{M}}(\rho) \cdot \bm{p}_{\mathcal{A}},
\end{eqnarray}
where
\begin{equation}
{\underline{M}}(\mathbf{\rho })\equiv \left( 
\begin{array}{cc}
\frac{4u^{2}}{\rho ^{4}}-\frac{2}{\rho ^{2}} & \frac{4uv}{\rho ^{4}} \\ 
\frac{4uv}{\rho ^{4}} & \frac{4v^{2}}{\rho ^{4}}-\frac{2}{\rho ^{2}}
\end{array}
\right)\label{mrho}
\end{equation}
in the two dimensional ($a$,$b$) coordinate system
(the dipole moment
in the $c$-direction is zero and hence we work with
$2\times 2$ instead of $3\times 3$ matrices). The microscopic electric field 
$\bm{E}'_{\rm{micr}}$ at the origin, excluding the
field induced by chain at the origin itself, is given by:
\begin{equation}
\bm{E}'_{\rm{micr}}(\vec{\rho}=0) = \bm{E}_{\rm{appl}}(\vec{\rho}=0)
 + {\underline{M}}_{\mathcal{A}}\cdot \bm{p}_{\mathcal{A}}+
{\underline{M}}_{\mathcal{B}}\cdot \bm{p}_{\mathcal{B}}, \label{Eyz}
\end{equation}
where
\begin{eqnarray}
{\underline{M}}_{\mathcal{A}} &\equiv& \lim_{k\rightarrow 0}
\sum_{\vec{\rho_i} \in {\mathcal{A}},\vec{\rho_i}\neq 0} {\underline{M}}(\vec{\rho_i})
\cos ku_i \label{ma},\\
{\underline{M}}_{\mathcal{B}} &\equiv& \lim_{k\rightarrow 0}\sum_{\vec{\rho_i} \in {\mathcal{B}}} {\underline{M}}(\vec{\rho_j})
\cos ku_j \label{mb},
\end{eqnarray}
These sums are evaluated in the next subsection.
Substitution of Eq.\ (\ref{Eyz}) in Eq.\ (\ref{pA}) and solving 
yields:
\begin{equation}
\tilde{\chi}_a \equiv p_{{\mathcal{A}},a}/E_{{\rm{appl}},a} = \left(
{\underline{\chi}}_{\mathcal{A}} \left[ {\underline{1}} -
{\underline{\chi}}_{\mathcal{A}}\cdot {\underline{M}}_{\mathcal{A}}-
{\underline{\chi}}_{\mathcal{B}}\cdot {\underline{M}}_{\mathcal{B}}
\right]^{-1} \right)_{aa},
\label{pyeq}
\end{equation}
with ${\underline{\chi}}_{\mathcal{A}}$ and ${\underline{\chi}}_{\mathcal{B}}$ as
defined in Eq.\ (\ref{chiB}).
Analogous to the derivation given by Jackson~\cite{jackson} for a point dipole, we can derive
the electric field of a line-dipole at $\vec{\rho}=0$:\cite{peter}
\begin{equation}
\bm{E}(\vec{\rho}) = \left({\underline{M}}(\vec{\rho}) - 2\pi\delta(\vec{\rho}){\underline{1}}\right)\cdot \bm{p}
\label{Erho}
\end{equation}
where the convention in Eq.\ (\ref{Erho}) is that the field within the line-dipole
at $\vec{\rho}=0$ is given by the term $-2\pi \delta (\vec{\rho}){\mathbf{p}}$ and the Cauchy principal value 
of the integral should be taken in integrals across the $1/\rho^2$ singularity at $\vec{\rho}=0$. 
The macroscopic field is given by the average over the microscopic field of Eq.\ (\ref{Eyz})
for general $\vec{\rho}$ including the chain at the origin.
Note that since, by symmetry, $p_{{\mathcal{A}},b} = - p_{{\mathcal{B}},b}$, the $b$ components do
not contribute to the macroscopic field. Also by symmetry, we have $p_{{\mathcal{A}},a} = 
p_{{\mathcal{B}},a} = p_a$. We then have for the macroscopic field $E_{a}(\vec{\rho})$ :
\begin{eqnarray}
E_{a}(\vec{\rho}) &=& E_{{\rm{appl}},a}+
 \lim_{k\rightarrow 0} \frac{1}{2A} {\mathcal{P}}\int_{\rm{unit}\mbox{ }cell} d^2\vec{\rho}
\sum_{\rho_k}M_{aa}(\vec{\rho}-\vec{\rho_k})\cos(ku_k)p_{a} - \frac{2\pi}{A}p_{a}\\
&=& E_{{\rm{appl}},a}+
 \lim_{k\rightarrow 0} \frac{1}{A} {\mathcal{P}}\int d^2\vec{\rho}
M_{aa}(\vec{\rho})\cos(ku)p_{a} - \frac{2\pi}{A}p_{a}\\
&=& E_{{\rm{appl}},a} -\frac{4\pi p_a}{A}.
\end{eqnarray}
Substituting
this result in Eq. (\ref{eqforeps}) and using the fact that $P_a=p_a/A$, we find Eq.\ (\ref{epsall}):
\begin{equation}
\varepsilon_{a} = \frac{1}{1 - {\frac{4\pi}{A} \tilde{\chi}_a}}.
\end{equation}
A similar result is obtained for $\varepsilon_b$.

\subsection{Evaluation of ${\underline{M}}_{\mathcal{A}}$ and ${\underline{M}}_{\mathcal{B}}$}

From the symmetry of Eqs.\ (\ref{mrho}), (\ref{ma}) and (\ref{mb}),
we see that ${\underline{M}}_{{\mathcal{A}},ab} = {\underline{M}}_{{\mathcal{A}},ba} = 0 =
{\underline{M}}_{{\mathcal{B}},ab} = {\underline{M}}_{{\mathcal{B}},ba} = 0$
and ${\underline{M}}_{{\mathcal{A}},aa}
 = - {\underline{M}}_{{\mathcal{A}},bb}$
and ${\underline{M}}_{{\mathcal{B}},aa} = - {\underline{M}}_{{\mathcal{B}},bb}$. 
This leaves us with only one element
of each matrix to be determined. Considering ${\underline{M}}_{\mathcal{A}}$
first, we split the summation of Eq.\ (\ref{ma}) into
two parts. For $\rho_i < R$ (with $R$ large) we perform the summation explicitly (taking $k=0$),
while for $\rho_i \geq R$ we replace the summation by an integral:
\begin{equation}
{\underline{M}}_{{\mathcal{A}},aa} =  \sum_{\vec{\rho_i} \in {\mathcal{A}} \mbox{ }
\vec{\rho_i} \neq 0 \mbox{ } \rho_i < R} {\underline{M}}_{aa}(\vec{\rho_i})
+ \lim_{k\rightarrow 0}\frac{1}{2A}
\int_R^{\infty}\!\!\rho d\rho\!\!\int_0^{2\pi}\!\!d\phi {\underline{M}}_{aa}(\rho)\cos(k\rho\cos\phi),
\end{equation}
which is exact in the limit $R \rightarrow \infty$.
The sum is evaluated numerically; its values is
$-0.009677$ $a_0^{-2}$ in the limit $R\rightarrow\infty$. The integral becomes
$-\pi/(2A)$ after {\em{first}} taking the limit $k \rightarrow 0$ and {\em{then}}
the limit $R \rightarrow \infty$.
We can calculate ${\underline{M}}_{{\mathcal{B}},aa}$ in a similar way. The sum yields $0.012035$ $a_0^{-2}$ and
the integral becomes again $-\pi/(2A)$. 
Therefore, ${\underline{M}}_{\mathcal{A}}$ and ${\underline{M}}_{\mathcal{B}}$ are:
\begin{eqnarray}
{\underline{M}}_{\mathcal{A}} &=& \left(\begin{array}{ll}-0.030068\mbox{ }a_0^{-2} & 0 \\\phantom{-}0&0.030068\mbox{ }a_0^{-2}\end{array}\right),\\
{\underline{M}}_{\mathcal{B}} &=& \left(\begin{array}{ll}-0.008357\mbox{ }a_0^{-2}& 0 \\ \phantom{-}0&0.008357\mbox{ }a_0^{-2}\end{array}\right).
\end{eqnarray}

\begin{table}
\begin{tabular}{l r @{.} l r @{.} l r @{.} l r @{.} l r @{.} l }
& \multicolumn{2}{c}{intra} & \multicolumn{6}{c}{intra+inter} & \multicolumn{2}{c}{experiment}  \\
$r_{\rm{inter}}$ (a.u.) & \multicolumn{2}{c}{} & 8&0 & 10&0 & 12&0 & \multicolumn{2}{c}{}  \\
\hline
$E_g$   & 3&59 &  2&32 & 2&49 & 2&69  & \multicolumn{2}{c}{} \\
$E_b (^1{\rm{B}}_{\rm{u}})$ & 1&85 & 0&64 & 0&76 & 0&86 & \multicolumn{2}{c}{} \\
$E_o$ & 1&74 & 1&68 & 1&73 & 1&73 & 1&8 \\
\multicolumn{7}{c} { } \\
$^3{\rm{B}}_{\rm{u}} \rightarrow \mbox{ }^1{\rm{B}}_{\rm{u}} $ & 0&51 & 0&34 & 0&39 & 0&45 & 0&45 
\\
$^1{\rm{B}}_{\rm{u}} \rightarrow \mbox{ }^1{\rm{A}}_{\rm{g}} $ & 0&89 & 0&45 & 0&53 & 0&58 & 0&55 
\\
\end{tabular}
\caption{Quasi-particle ($E_g$) and optical ($E_o$) gaps and binding energies ($E_b$),
for the cases `intra', using intrachain screening only (isolated chain), and `intra+inter',
using both intra-and interchain screening (bulk) for three different values of the cut-off 
distance $r_{\rm{inter}}$ (see Eq.\ (\protect\ref{weps})). Exciton transition energies are also 
listed. Experimental data from Ref.\ 
\protect\onlinecite{sakurai}. All data in eV.}
\label{excagain}
\end{table}
\begin{table}
\begin{tabular}{c r@{.}l c r@{.}l } 
\multicolumn{3}{c}{Polarizabilities} & \multicolumn{3}{c}{Dielectric constants}\\
Direction  $j$ & \multicolumn{2}{c}{$\chi_{i}(\omega = 0)$} & Crystal Axis $\gamma$ &
  \multicolumn{2}{c}{$\varepsilon_{\gamma}(\omega = 0)$} \\
\hline
1 & 60&4 & $c$ & 10&8\\
2 & 16&3 & $b$ & 3&3\\
3 & 8&1 &  $a$ & 2&6\\
\end{tabular}
\caption{The zero-frequency polarizabilities
 of the single chain per unit length (in $a_0^2$)
and the
dielectric constants of the bulk along the principal axes of the chain and the
crystal.}
\label{poleps}
\end{table}


\begin{thebibliography}{99}

\bibitem{ppv}
J.H.\ Burroughes, D.D.C.\ Bradley, A.R.\ Brown, R.N.\ Marks, K. Mackay,
R.H.\ Friend, P.L.\ Bum, A.B.\ Holmes, Nature {\bf{347}}, 359 (1990).

\bibitem{bredasheeger}
J.-L. Br\'edas, J. Cornil, A.J. Heeger, Adv. Mat. {\bf{8}}, 447 (1996).

\bibitem{onetenthorless}
T.W.\ Hagler, K.\ Pakbaz, A.J.\ Heeger, Phys.\ Rev.\ B {\bf{49}}, 10968 (1994).

\bibitem{intermediate}
S.F.\ Alvarado, P.F.\ Seidler, D.G.\ Lidzey, D.D.C.\ Bradley, Phys.\ Rev.\ Lett.\ {\bf{81}}, 1082 (1998).

\bibitem{oneev}
M.\ Chandross, S.\ Mazumdar, S.\ Jeglinski, X.\ Wei, Z.V.\ Vardeny, E.W.\ Kwock, T.M.\ Miller,
Phys.\ Rev. {\bf{B}} 50, 14702 (1994).

\bibitem{melppp}
S.\ Barth, H.\ B\"assler, U.\ Scherf, K.\ M\"ullen, Chem.\ Phys.\ Lett.\ {\bf{288}}, 147 (1998).

\bibitem{wohl}
M.\ Wohlgenannt, W.\ Graupner, G.\ Leising, Z.V.\ Vardeny, Phys.\ Rev. B {\bf{60}}, 5321 (1999).

\bibitem{vc}
P.\ Vogl, D.K.\ Campbell, Phys.\ Rev.\ B {\bf{41}}, 12797 (1990).

\bibitem{bkc}
G. Brocks, P.J. Kelly, R. Car, Synth.\ Met.\ {\bf{55-57}}, 4243 (1993).

\bibitem{gdc}
P.\ Gomes da Costa, R.G.\ Dandrea, E.M.\ Conwell, Phys.\ Rev.\ B {\bf{47}}, 1800 (1993).

\bibitem{abd}
C.\ Ambrosch-Draxl, J.A.\ Majewski, P.\ Vogl, G.\ Leising, Phys.\ Rev.\ B {\bf{51}}, 9668 (1995).

\bibitem{godbydeltaverhaal}
R.W.\ Godby, M.\ Schl\"uter, L.J.\ Sham, Phys.\ Rev.\ B {\bf{37}}, 10159 (1988).

\bibitem{gw} L. Hedin, Phys.\ Rev.\ {\bf 139}, A796 (1965); for a recent review see:
F. Aryasetiawan and O. Gunnarsson, Rep.\ Prog.\ Phys.\ {\bf 61}, 237 (1998).

\bibitem{ethridge}
E.C.\ Ethridge, J.L.\ Fry, M.\ Zaider, Phys.\ Rev.\ {\bf{B}} 53, 3662 (1996).

\bibitem{oeicp}
M. Rohlfing and S.G. Louie, Phys.\ Rev.\ Lett.\ {\bf{82}}, 1959 (1999).

\bibitem{alvpriv}
S.F.\ Alvarado (private communication).

\bibitem{onszelf}
J.W. van der Horst, P.A. Bobbert, M.A.J. Michels, G. Brocks, P.J. Kelly,
Phys. Rev. Lett. {\bf{83}}, 4413 (1999).

\bibitem{sakurai}
K. Sakurai, H. Tachibana, N. Shiga, C. Terakura, M. Matsumoto and Y. Tokura,
Phys. Rev. B {\bf{56}}, 9552 (1997).

\bibitem{herman}
H.J. de Groot, P.A. Bobbert, and W. van Haeringen, Phys.\ Rev.\ {\bf{B}} 52, 11000 (1995).

\bibitem{brsawa}
J.\ van den Brink and G.A.\ Sawatzky, in {\em{Electronic Properties of Novel Materials-Progress in 
Molecular Nanostructures}}, edited by H.\ Kuzmany, M.\ Mehring, and S.\ Roth,
AIP Conf.\ Proc.\ No.\ 442 (AIP, New York, 1998), p.\ 152.

\bibitem{schulz2}
H.J.\ Schulz, Phys.\ Rev.\ Lett.\ {\bf{71}}, 1864 (1993), 
and references therein.

\bibitem{ben}
L.X.\ Benedict, R.B.\ Bohn, and E.L.\ Shirley, Phys. Rev. B {\bf{57}} R9385 (1998),
Phys. Rev. Lett. {\bf{80}}, 4514 (1998); L.X.\ Benedict and E.L.\ Shirley, Phys. Rev.
B {\bf{59}}, 5441 (1999).

\bibitem{ar} 
S. Albrecht, L. Reining, R. Del Sole and G. Onida.,
Phys.\ Rev.\ Lett.\ {\bf{80}}, 4510 (1998).

\bibitem{rl}
M. Rohlfing and S.G. Louie, Phys.\ Rev.\ Lett.\ {\bf{81}}, 2312 (1998).

\bibitem{shamrice}
L. Sham and T.M. Rice, Phys.\ Rev.\ {\bf{144}}, 708 (1966).

\bibitem{strinati}
G.\ Strinati, Phys.\ Rev.\ B {\bf{29}}, 5718 (1984).

\bibitem{polaron}
G.\ Brocks, Synth.\ Met.\ {\bf{102}}, 914 (1999).

\bibitem{tripexc}
G.\ Brocks, unpublished. With DFT the ground state of a given symmetry can be obtained.
The triplet ground state corresponds to the lowest triplet exciton.

\bibitem{xtal}
S. Br\"uckner and W. Porzio, Makromol. Chem. {\bf{189}}, 961 (1988).

\bibitem{mo}
Z.\ Mo, K.-B.\ Lee, Y.B.\ Moon, M.\ Kobayashi, A.J.\ Heeger, F.\ Wudl, Macromol. {\bf{18}}, 1972 (1985).

\bibitem{spacetime}
H.N. Rojas, R.W. Godby, and R.J. Needs, Phys. Rev. Lett. {\bf{74}}, 1827 (1995);
for a more detailed account see:
M. M. Rieger, L. Steinbeck, I.D. White, H.N. Rojas and R.W. Godby, Comp.
Phys. Comm. {\bf{117}}, 211 (1999).

\bibitem{msform}
X.\ Blase, A. Rubio, S.G.\ Louie, M.L.\ Cohen, Phys.\ Rev.\ B {\bf{52}}, R2225 (1995).

\bibitem{bechstedt}
F.\ Bechstedt, K.\ Tenelsen, B.\ Adolph, R.\ Del Sole,
Phys.\ Rev.\ Lett.\ {\bf{78}}, 1528 (1997).

\bibitem{ll}
L.\ Landau and E.\ Lifschitz, {\em{Electrodynamics of Continuous Media}} (Pergamon,
Oxford, 1960), p.\ 61-62.

\bibitem{kobi}
M. Kobayashi, N. Colaneri, M. Boysel, F. Wudl and A.J. Heeger,
J.\ Chem.\ Phys.\ {\bf{82}}, 5717 (1985).

\bibitem{bwsh}
E.L.\ Shirley, Phys.\ Rev.\ B {\bf{58}}, 9579 (1999)

\bibitem{absvslum}
R.\ Kersting, U.\ Lemmer, R.F.\ Mahrt, K.\ Leo, H.\ Kurz, H.\ B\"assler, E.O.\ G\"obel,
Phys.\ Rev.\ Lett.\ {\bf{70}}, 3820.

\bibitem{gdc2}
P.\ Gomes da Costa and E.M. Conwell, Phys.\ Rev.\ B {\bf{48}}, R1993 (1993).

\bibitem{jackson}
J.D.\ Jackson, {\em{Classical Electrodynamics (2nd edition)}} (Wiley, New York, 1975), 
p.\ 139-141.

\bibitem{peter}
P.H.L.\ de Jong, M.Sc. Thesis, Technische Universiteit Eindhoven, 1999.

\end{thebibliography}
\end{document}